\documentstyle[12pt]{article}
\catcode`\@=11

\@addtoreset{equation}{section}
\def\theequation{\arabic{section}.\arabic{equation}}
\catcode`\@=11

\def\section{\@startsection{section}{1}{\z@}{3.5ex plus 1ex minus
   .2ex}{2.3ex plus .2ex}{\large\bf}}
\def\eqnarray{\stepcounter{equation}\let\@currentlabel=\theequation%
\global\@eqnswtrue
    \global\@eqcnt\z@\tabskip\@centering\let\\=\@eqncr
    $$\halign to \displaywidth\bgroup\@eqnsel\hskip\@centering
      $\displaystyle\tabskip\z@{##}$&\global\@eqcnt\@ne
       \hfil${{}##{}}$\hfil
      &\global\@eqcnt\tw@ $\displaystyle\tabskip\z@{##}$\hfil
       \tabskip\@centering&\llap{##}\tabskip\z@\cr}
\def\lefteqn#1{\hbox to 4\arraycolsep{$\displaystyle #1$\hss}}

\long\def\@makefntext#1{\parindent 0cm\noindent
\hbox to 1em{\hss$^{\@thefnmark}$}#1}
\def\rref#1{(\ref{#1})}
\newcommand{\beq}{\begin{equation}}
\newcommand{\eeq}{\end{equation}}
\newcommand{\vol}{{\tilde v}}
\begin{document}
%%%%%%%%%%%%%%%%%%%%%%%%%%%%%%%%%%%%%%%%%%%%%%%%%%%%%%%%%%%%%%%%%%%%%%%%%%%
%     C I T E . S T Y
%     compressed lists of numerical citations: [11-16]
%     see also OVERCITE.STY and DRFTCITE.STY
%
%     Copyright (C) 1989-1992 by Donald Arseneau
%     These macros may be freely transmitted, reproduced, or modified for
%     non-commercial purposes provided that this notice is left intact.
%
%
%  \@citen contains the code that parses the list of names, ignoring
%  spaces after commas, writes the aux file \citation, and formats the
%  number list.  \citen can be used by itself to give citation numbers
%  without the other formatting; e.g., "See also ref.~\citen{junk}."
%
\def\citen#1{%
\edef\@tempa{\@ignspaftercomma,#1, \@end, }% ignore spaces in parameter list
\edef\@tempa{\expandafter\@ignendcommas\@tempa\@end}%
\if@filesw \immediate \write \@auxout {\string \citation {\@tempa}}\fi
\@tempcntb\m@ne \let\@h@ld\relax \let\@citea\@empty
\@for \@citeb:=\@tempa\do {\@cmpresscites}%
\@h@ld}
%
% for ignoring spaces in the input:
\def\@ignspaftercomma#1, {\ifx\@end#1\@empty\else
   #1,\expandafter\@ignspaftercomma\fi}
\def\@ignendcommas,#1,\@end{#1}
%
% For each citation, check if it is defined, if it is a number, and
% if it is a consecutive number that can be represented like 3-7.
%
\def\@cmpresscites{%
 \expandafter\let \expandafter\@B@citeB \csname b@\@citeb \endcsname
 \ifx\@B@citeB\relax % undefined
    \@h@ld\@citea\@tempcntb\m@ne{\bf ?}%
    \@warning {Citation `\@citeb ' on page \thepage \space undefined}%
 \else%  defined
    \@tempcnta\@tempcntb \advance\@tempcnta\@ne
    \setbox\z@\hbox\bgroup % check if citation is a number:
    \ifnum\z@<0\@B@citeB \relax
       \egroup \@tempcntb\@B@citeB \relax
       \else \egroup \@tempcntb\m@ne \fi
    \ifnum\@tempcnta=\@tempcntb % Number follows previous--hold on to it
       \ifx\@h@ld\relax % first pair of successives
          \edef \@h@ld{\@citea\@B@citeB}%
       \else % compressible list of successives
%         % use \hbox to avoid easy \exhyphenpenalty breaks
          \edef\@h@ld{\hbox{--}\penalty\@highpenalty \@B@citeB}%
       \fi
    \else   %  non-successor--dump what's held and do this one
       \@h@ld \@citea \@B@citeB \let\@h@ld\relax
 \fi\fi%
 \let\@citea\@citepunct
}
%
%%    To put space after the comma, use:
\def\@citepunct{,\penalty\@highpenalty\hskip.13em plus.1em minus.1em}%
%%    For no space after comma, use:
%% \def\@citepunct{,\penalty\@highpenalty}%
%%
%
%  Make \@citex refer to \citen:
%
\def\@citex[#1]#2{\@cite{\citen{#2}}{#1}}%
%
%  Replacement for \@cite.  Give one normal space before the citation,
%  set high penalties for linebreaks,
%
\def\@cite#1#2{\leavevmode\unskip
  \ifnum\lastpenalty=\z@ \penalty\@highpenalty \fi % highpenalty before
  \ [{\multiply\@highpenalty 3 #1% % triple-highpenalties within list
      \if@tempswa,\penalty\@highpenalty\ #2\fi % and before note.
    }]\spacefactor\@m}
\let\nocitecount\relax  % in case \nocitecount was used for drftcite
%
%%%%%%%%%%%%%%%%%%%%%%%%%%%%%%%%%%%%%%%%%%%%%%%%%%%%%%%%%%%%%%%%%%%%%%%%%%
\begin{titlepage}
\vspace{.5in}
\begin{flushright}
UCD-97-21\\
October 1997\\
revised January 1998\\
gr-qc/9710114\\
\end{flushright}
\vspace{.5in}
\begin{center}
{\Large\bf
 Dominant Topologies\\[1ex] in Euclidean Quantum Gravity}\\
\vspace{.4in}
{S.~C{\sc arlip}\footnote{\it email: carlip@dirac.ucdavis.edu}\\
       {\small\it Department of Physics}\\
       {\small\it University of California}\\
       {\small\it Davis, CA 95616}\\{\small\it USA}}
\end{center}

\vspace{.5in}
\begin{center}
{\large\bf Abstract}
\end{center}
\begin{center}
\begin{minipage}{4.75in}
{\small
The dominant topologies in the Euclidean path integral for quantum
gravity differ sharply according on the sign of the cosmological
constant.  For $\Lambda>0$, saddle points can occur only for topologies
with vanishing first Betti number and finite fundamental group.  For
$\Lambda<0$, on the other hand, the path integral is dominated by
topologies with extremely complicated fundamental groups; while
the contribution of each individual manifold is strongly suppressed,
the ``density of topologies'' grows fast enough to overwhelm this
suppression.  The value $\Lambda=0$ is thus a sort of boundary between
phases in the sum over topologies.  I discuss some implications for
the cosmological constant problem and the Hartle-Hawking wave function.
}
\end{minipage}
\end{center}
\end{titlepage}
\addtocounter{footnote}{-1}

It has been forty years since Wheeler first suggested that the topology
of spacetime might be subject to quantum fluctuations \cite{Wheeler}.
We do not yet know whether the resulting picture of ``spacetime foam''
correctly describes the universe, but the potential implications are
clearly important: for example, fluctuations of topology are a key element
in Coleman's proposed wormhole/baby universe solution to the cosmological
constant problem \cite{Coleman}.  If such fluctuations occur only at the
Planck scale, a full-fledged quantum theory of gravity may be necessary
to understand their effect.  If they can occur at larger scales, however,
it may be possible to treat the standard Einstein action as an effective
field theory \cite{effect} from which we can draw useful conclusions.

To understand the quantum mechanics of spacetime topology, one needs a
formalism in which spacetime is treated as a unified entity.  Canonical
quantum gravity may allow us to investigate changes in spatial topology,
but a path integral approach seems more natural if we are interested in
the topology of spacetime as a whole.  In particular, much of the work
on this subject (see, for example, \cite{Hawking1,Hawking2,Hawking3,%
Gibbons1,Gibbons2,Eguchi,Page,Christensen,Hartle,Hosoya,Carlip1})
has been based on path integral techniques in Euclidean quantum gravity,
that is, general relativity ``Wick rotated'' to Riemannian (positive
definite) metrics.

If the Einstein action is treated as part of an effective field theory
for distances larger than the Planck length, one should not worry too
much about higher-loop corrections, which will be suppressed by powers of
the Planck mass.  It is thus sensible to treat the path integral in a
saddle point approximation.  The purpose of this article is to describe
some features of saddle points and to discuss possible implications for
spacetime foam.  Some of the results presented here are old, but are
not widely known among physicists; others are new.  A brief report on
results for $\Lambda<0$ has appeared in reference \cite{Carlip2}.

\section{The Euclidean Path Integral}

In the Euclidean path integral approach to quantum gravity, the simplest
quantity to compute is the partition function \cite{Hawking1}, which can
be written formally as a path integral
\beq
Z[\Lambda] = \sum_M \int [dg] \exp\{-I_E[g]\} .
\label{a1}
\eeq
The Euclidean action $I_E$ in equation \rref{a1} is
\beq
I_E[g] = - {1\over16\pi L_P{}^2} \int_M (R-2\Lambda) \sqrt{g}\,d^4x,
\label{a2}
\eeq
where $g$ is a Riemannian metric on the manifold $M$ and $L_P$ denotes
the Planck length.  The sum in \rref{a1} is a ``sum over topologies,''
that is, a sum over topologically distinct manifolds $M$.  It should be
noted from the outset that the meaning of such a sum is not entirely clear.
Four-manifolds are not classifiable---that is, there is no algorithm that
can determine whether two arbitrary four-manifolds are homeomorphic---and
the sum over topologies may yield a number that is noncomputable in the
sense of Turing.\footnote{It is not known whether this is an inherent
characteristic of the partition function or merely a problem with
the particular representation \rref{a1}.  The existence of a
``noncomputable'' expression for a number is not sufficient to show
that the number itself is noncomputable; for example, a sequence of
noncomputable numbers can have a computable limit \cite{Geroch}.}
Geroch and Hartle have discussed the implications of noncomputability
for physics \cite{Geroch}, and argue that it need not be a disaster:
even if a quantity like the partition function is not computable, one
may be able to obtain approximations to any desired degree of accuracy.

In principle, one might extend the sum \rref{a1} to objects other than
manifolds---for instance, pseudomanifolds \cite{Hartle} or ``conifolds''
\cite{Schleich1,Schleich2}.  Conifolds, in particular, occur at the
boundaries of moduli spaces of Einstein metrics, and are classifiable,
thus allowing us to evade the problem of noncomputability discussed above.
One might also introduce relative phases between terms in the sum.  In
simpler systems, such phases are restricted by the requirement that
amplitudes behave correctly under composition \cite{Laidlaw}, but little
in known about the case of gravity.  For simplicity, I will largely
ignore such generalizations, which are unlikely to affect the main
conclusions of this paper.

An extremum of the action \rref{a2} ia an Einstein metric, that is, a
metric for which
\beq
R_{\mu\nu} = \Lambda g_{\mu\nu} .
\label{a3a}
\eeq
The classical action for such a metric is
\beq
\bar I_E(M,g) = -{\Lambda\over8\pi L_P{}^2}{\mathit{Vol}}(M,g) .
\label{a3}
\eeq
This expression is slightly misleading, however, since the volume of $M$
depends implicitly on the cosmological constant.  We can isolate this
dependence by rescaling the metric to set the scalar curvature to $\pm12$.
(The factor of $12$ is conventional; four-manifolds of constant curvature
$\pm1$ have scalar curvature $\pm12$.)  To do so, we define
\beq
g_{\mu\nu} = {3\over|\Lambda|} {\tilde g}_{\mu\nu} ,
\label{a4}
\eeq
where the rescaled metric $\tilde g$ satisfies \rref{a3a} with $\Lambda
=\pm3$.  The action \rref{a3} is then
\beq
\bar I_E(M,g) = - {9\over8\pi\Lambda L_P{}^2}\vol(M,g) ,
\label{a5}
\eeq
where the normalized volume $\vol(M,g)$ is the volume with respect to
$\tilde g$, and the only dependence on $\Lambda$ now resides in the
overall $1/\Lambda$ factor.

The normalized volume $\vol$ is clearly a geometric quantity, but it
is also, in a sense, topological: the set of normalized volumes of
Einstein metrics on a manifold $M$ characterizes the topology of $M$.
In particular, for $\Lambda<0$ there is no known example of a manifold
that admits two Einstein metrics with different values of $\vol$
\cite{Besse}.  Roughly speaking, $\vol(M,g)$ measures the topological
complexity of $M$; for a four-manifold with a constant curvature metric
$g_0$, for instance, $\vol(M,g_0) = 4\pi^2\chi(M)/3$, where $\chi$ is
the Euler characteristic.  The normalized volume is also closely related
to the ``minimal volume,'' a topological invariant defined as
\beq
{\mathit{minvol}}(M)
  = {\mathit{inf}} \{ {\mathit{Vol}}(M,g) | \,|K_g|\le 1 \} .
\label{a5a}
\eeq
Here $K_g$ is the sectional curvature, that is, the quadratic form
\beq
K_g(v,w) = {R_{\mu\nu\rho\sigma}v^\mu w^\nu v^\rho w^\sigma \over
   (v_\mu v^\mu)(w_\nu w^\nu) - (v_\mu w^\mu)^2 }
\label{a5b}
\eeq
and the infimum is over Riemannian metrics on $M$.  Gromov has
conjectured that for any manifold $M$ that admits a hyperbolic metric
$g_0$, ${\mathit{minvol}}(M) = \vol(M,g_0)$ \cite{Gromov}; a local
version of this conjecture has recently been proven \cite{Besson2}.

In the saddle point approximation, the partition function \rref{a1} is now
\beq
Z[\Lambda] = \sum_{(M,g)} \Delta_{(M,g)}\exp
  \left\{ {9\over8\pi\Lambda L_P{}^2}\vol(M,g) \right\} ,
\label{a6}
\eeq
summed over pairs $(M,g)$ of four-manifolds with Einstein metrics.  The
prefactors $\Delta_{(M,g)}$ are combinations of Faddeev-Popov determinants
coming from gauge-fixing and Van Vleck-Morette determinants coming from
small fluctuations around the extrema.  Their precise values are not known,
but the dependence of $\Delta_{(M,g)}$ on $\Lambda$ can be computed from
the trace anomaly \cite{Christensen}: up to possible polynomial corrections
coming from zero-modes,
\beq
\Delta_{(M,g)} \sim \Lambda^{-\gamma/2} , \qquad
\gamma = {106\over45}\chi(M) - {261\over40\pi^2}\vol(M,g) .
\label{a7}
\eeq
For our purposes, the crucial observation is that $\Delta_{(M,g)}$ is no
more than exponential in $\vol$.

It is useful to rewrite the sum over topologies in equation \rref{a6} as
a sum over normalized volumes,
\beq
Z[\Lambda] = \sum_\vol \rho(\vol) \exp
  \left\{ {9\over8\pi\Lambda L_P{}^2}\vol \right\} .
\label{a8}
\eeq
The factor $\rho(\vol)$ is a ``density of topologies'' that counts the
number of Einstein manifolds (weighted by $\Delta_{(M,g)}$) with a given
value of $\vol$.  For $\Lambda>0$, the exponent in \rref{a8} picks out
the manifold with the largest normalized volume, the four-sphere $S^4$.
For $\Lambda<0$, manifolds with large normalized volumes---hyperbolic
manifolds with large Euler characteristics, for example---are exponentially
suppressed.  In either case, however, there can be competition between
the exponential factor (or ``action'') and the density of topologies (or
``entropy'').  In three spacetime dimensions, it is known that the
``entropy'' can, in fact, dominate the ``action'' \cite{Carlip1,Carlip3};
one goal of this paper is to see whether the same is true in four dimensions.

Before proceeding further, let us note that not every manifold admits
an Einstein metric with any value of the cosmological constant.  In fact,
as simple a manifold as the connected sum of two tori, $T^4\#T^4$,
admits no Einstein metric \cite{Besse}.  As Hitchin and Thorpe have shown
\cite{Hitchin,Thorpe}, in order for a manifold $M$ to admit an Einstein
metric, its Euler characteristic and signature must obey the inequality
\beq
\chi(M) \ge {3\over2}|\tau(M)| .
\label{a9}
\eeq
This condition is necessary, but not sufficient: LeBrun \cite{LeBrun}
and Sambusetti \cite{Sambusetti} have constructed infinitely many compact
four-manifolds that satisfy the Hitchin-Thorpe inequality but admit no
Einstein metric.  My philosophy will be that such manifolds can be
ignored in the sum over topologies; they are relevant only in higher-loop
approximations, which are important at scales at which the Einstein action
no longer makes sense as an effective action and the whole approach to
quantum gravity must be reconsidered.

\section{Positive Cosmological Constant}

Let us begin by examining the case $\Lambda>0$.  From equation \rref{a8},
the largest individual contribution to the partition function will come
from the manifold that admits an Einstein metric with the largest value
of $\vol$.\footnote{The Euclidean path integral may be misleading in
this case, however.  Starting with the canonical formalism for simple
reparametrization-invariant systems, Marolf has argued that the correct
contribution for a Euclidean instanton is $\exp\{-|{\bar I}_E|\}$
\cite{Marolf}.  If this is the case, the sign of the exponent in \rref{a8}
should be changed when $\Lambda>0$, and large values of $\vol$ will
be suppressed.}  In four dimensions, this is the four-sphere $S^4$
\cite{Bishop}, which has a normalized volume
\beq
\vol(S^4) = {8\pi^2\over3} .
\label{b1}
\eeq
Relatively few other examples of manifolds admitting Einstein metrics
with $\Lambda>0$ are known explicitly.  Examples include ${\mathbf{CP}}^2$,
$S^2\times S^2$, and the Page metric on an $S^2$ bundle over $S^2$
\cite{Besse,GibHawk}.  Tian and Yau have also found a finite-dimensional
moduli space of Einstein metrics with $\Lambda>0$ on the manifolds
$${\mathbf{CP}}^2\# k\overline{\mathbf{CP}}^2 $$
for $5\le k\le 8$ \cite{Tian}.  (For a general discussion of moduli spaces
of Einstein metrics, see \cite{Besse,Anderson}.)

To date, very little is understood about the behavior of the ``density
of topologies'' $\rho(\vol)$ for $\Lambda>0$, and not much can be said
about the sum \rref{a8}.  We can, however, make some surprisingly strong
statements about topologies that do {\em not\/} appear in the partition
function in the saddle point approximation.  In particular, Myers has
shown that any complete Einstein manifold with a positive cosmological
constant necessarily has a finite fundamental group \cite{Myers}: there
are no Euclidean wormholes with $\Lambda>0$.  This fact is instrumental
in proving the ``no multiple birth'' theorem for the Hartle-Hawking wave
function with $\Lambda>0$ \cite{Hartle3,Gibbons2}.

The full proof of Myers' theorem is quite complicated, but a weaker
version is rather straightforward: it is easy to show that a closed
Einstein manifold with a positive cosmological constant must have
vanishing first Betti number \cite{Bochner,Berard}.  Indeed, suppose
the cohomology $H^1(M,\mathbf{R})$ is nontrivial.  Any element of this
cohomology can be represented by a harmonic one-form $\omega$, that
is, a one-form satisfying
\begin{eqnarray}
&&\nabla_\mu\omega_\nu - \nabla_\nu\omega_\mu = 0, \label{b2} \\
&&\nabla_\mu\omega^\mu = 0 \label{b3} .
\end{eqnarray}
We can now differentiate \rref{b2} and use \rref{b3} to obtain
\beq
\nabla^\mu\nabla_\mu\omega_\nu = \nabla^\mu\nabla_\nu\omega_\mu
   = [\nabla^\mu,\nabla_\nu]\omega_\mu = R_\nu{}^\rho\omega_\rho
   = \Lambda \omega_\nu .
\label{b4}
\eeq
Contracting with $\omega^\nu$ and integrating over $M$, we see that
\begin{eqnarray}
0 &=& \int_M \left[ \Lambda\omega^\nu\omega_\nu
    - \omega^\nu\nabla^\mu\nabla_\mu\omega_\nu \right]\sqrt{g}\,d^4x
    \nonumber\\
  &=& \int_M \left[ \Lambda\omega^\nu\omega_\nu
    + (\nabla^\mu\omega^\nu)(\nabla_\mu\omega_\nu) \right]\sqrt{g}\,d^4x .
\label{b5}
\end{eqnarray}
But both terms in the last integrand are nonnegative, so \rref{b5}
implies that $\omega_\mu=0$, and thus $H^1(M,{\mathbf{R}})=0$.

Note that this proof does not really require that we have an Einstein
metric: it is enough to demand that the Ricci tensor in \rref{b4} be
positive everywhere.  Generalizations have been found for manifolds in
which the Ricci tensor is ``mostly'' positive, with appropriate restrictions
on regions of negative curvature \cite{Elworthy,Wu,Rosenberg}.  It is
tempting to speculate that these results are Euclidean versions of the
well-known fact that traversible Lorentzian wormholes require exotic
matter \cite{Visser}.

\section{Negative Cosmological Constant \label{sec1}}

We now turn to the case $\Lambda<0$.  Observe first that the dominant
contributions to the partition function may differ sharply depending
on the sign of $\Lambda$.  As we have seen, the manifolds that are
important when $\Lambda>0$ have relatively simple fundamental groups.
For $\Lambda<0$, on the other hand, a contribution---a very large
contribution, as we shall see below---comes from hyperbolic manifolds,
that is, manifolds with constant negative curvature.  Hyperbolic manifolds
in four dimensions are obtained as quotients $M\approx{\mathbf{H}}^4/\pi_1$
of hyperbolic four-space ${\mathbf{H}}^4$; they typically have very
complicated fundamental groups, and can have arbitrarily large first
Betti numbers \cite{Lubotzky2}.  Such manifolds make no contribution to
the $\Lambda>0$ partition function in the saddle point approximation.
Indeed, it has been shown that if a four-manifold $M$ admits a hyperbolic
metric, it is the {\em only\/} Einstein metric on $M$ \cite{Besson}.

As in the case of positive cosmological constant, we have nothing like
a complete classification of Einstein manifolds with $\Lambda<0$, but we
know a number of interesting examples.  In addition to hyperbolic manifolds
\cite{Ratcliffe}, these include product manifolds $\Sigma_1\times\Sigma_2$,
where $\Sigma_1$ and $\Sigma_2$ are surfaces of genus $h_1,h_2>1$
\cite{Gibbons3}, and compact complex manifolds with negative first Chern
class, which always admit K{\"a}hler-Einstein metrics with $\Lambda<0$
\cite{Aubin,Yau,Besse}.  In the latter two examples, one typically finds
a whole moduli spaces of metrics, with $\vol$ constant on the moduli space.
When this occurs, the prefactor $\Delta_M$ in \rref{a6} will include the
volume of the moduli space, and the density of topologies $\rho(\vol)$
should incorporate this factor.

In contrast to the $\Lambda>0$ case, we can now say something useful
about the function $\rho(\vol)$: while its complete behavior is not
understood, it may be shown that $\rho(\vol)$ increases at least
factorially with $\vol$.  Indeed, even if we restrict our attention
to the special case of hyperbolic manifolds, $\rho(\vol)$ still exhibits
at least factorial growth.  This means that the ``entropy'' in the sum
\rref{a8} dominates the action, and the partition function receives large
contributions from arbitrarily complicated topologies.

The following proof was explained to me by Lubotzky \cite{Lubotzky1}.
Observe first that if we limit our attention to hyperbolic four-manifolds,
the number of manifolds with normalized volumes $\vol(M,g)<V$ is finite for
any finite $V$ \cite{Wang}.  We thus need not worry about moduli spaces;
it suffices to simply count manifolds.  We begin with a hyperbolic ``seed
manifold'' $M$ with fundamental group $G$, metric $g$, and normalized
volume $\vol(M,g) = \vol_0$.  Any subgroup $G_1\subset G$ determines a
covering manifold $M_1$ of $M$ with fundamental group $G_1$, and $M_1$
inherits a hyperbolic metric $g_1$ from the metric $g$ on $M$.  Moreover,
if $G_1$ has index $n$ in $G$, then $M_1$ is an $n$-fold cover of $M$,
and $\vol(M_1,g_1) = n\vol_0$.\footnote{A subgroup $G_1$ has index $n$ in
$G$ if the number of distinct cosets $G_1g\ (g\in G)$ is $n$.}  Hence if
we can count the number of index-$n$ subgroups of $G$, and if we can avoid
double-counting isometric covering spaces, we can obtain a lower limit on
the number of hyperbolic manifolds with normalized volumes $n\vol_0$.

In reference \cite{Lubotzky2}, Lubotzky demonstrates the existence of a
hyperbolic manifold $M$ whose fundamental group $G$ maps homomorphically
onto a free group $F_r$ of rank $r>1$.  This result is useful because
the subgroup growth of free groups is well-understood \cite{Hall}: for
large $n$, the number of index-$n$ subgroups of a free group of rank
$r$ grows as
\beq
N(n,r) \sim n(n!)^{r-1} .
\label{c1}
\eeq
The existence of a surjective homomorphism $G\rightarrow F_r$ guarantees
that any index-$n$ subgroup of $F_r$ determines an index-$n$ subgroup of
$G$, so \rref{c1} gives a lower limit for the number of subgroups of $G$.

This is not yet the whole story, however.  Different subgroups $G_1,G_2
\subset G$ may sometimes give isometric covering spaces of $M$.  This
will occur if $G_1$ and $G_2$ are conjugate in $\hbox{SO}(4,1)$, the group
of isometries of ${\mathbf{H}}^4$; that is, $G_2 = g^{-1}G_1g$ for some
$g\in\hbox{SO}(4,1)$.  Fortunately, this condition can be simplified: it
may be shown that $G_1$ and $G_2$ must be conjugate in the commensurability
group $\hbox{Comm}(G)$ of $G$.\footnote{Two subgroups $G$ and $G'$ of a
group $\tilde G$ are commensurable if their intersection $G\cap G'$ has
finite index in each of them.  The commensurability group $\hbox{Comm}(G)$
of $G\subset{\tilde G}$ is the group $\{g\in{\tilde G}: g^{-1}Gg\ \hbox{is
commensurable with}\ G\}$.  In the case under consideration here, the
fundamental group $G$ is being viewed as a subgroup of $\hbox{SO}(4,1)$.
That $g\in\hbox{Comm}(G)$ then follows from the observation that
$g^{-1}Gg\cap G\supset G_2$, which has finite index in $G$.}  By a theorem
of Margulis, if $G$ is nonarithmetic, $\hbox{Comm}(G)$ is a finite extension
of $G$ \cite{Zimmer}.  Since Lubotzky showed in \cite{Lubotzky2} that the
``seed manifold'' $M$ could be chosen to have a nonarithmetic fundamental
group, we can concentrate on this case.

Suppose first that $G_1$ and $G_2$ are conjugate in $G$, that is, $G_2 =
g^{-1}G_1g$ for some $g\in G$.  Since $G_1$ is an index-$n$ subgroup of
$G$, there is a set $X$ of $n$ elements of $G$ such that $g = hx$ for
some $x\in X,\ h\in G_1$.  Hence $G_2 = x^{-1}G_1x,\ x\in X$, which means
there can be at most $n$ subgroups conjugate to $G_1$.  Now, $G_1$ and
$G_2$ may actually be conjugate in the larger group $\hbox{Comm}(G)$.  But
since $\hbox{Comm}(G)$ is a finite extension of $G$, a similar argument
shows that there can be at most $kn$ conjugate index-$n$ subgroups, where
$k$ is the index of $G$ in $\hbox{Comm}(G)$.  The estimate \rref{c1} thus
overcounts covering spaces by at most a factor of $kn$.

Combining these results, we obtain a bound
\beq
\rho(n\vol_0) \ge \hbox{const.}\ (n!)^{r-1}
\sim \exp \bigl\{ (r-1)n\log n\bigr\}
\label{c2}
\eeq
for the density of topologies.  It should be clear that this is merely a
lower bound---we have considered only hyperbolic manifolds, and only a
small subset of hyperbolic manifolds at that.  But this result is already
sufficient to demonstrate the superexponential growth of $\rho(\vol)$
with $\vol$, thus showing that the ``entropy'' dominates the ``action'' in
the sum \rref{a8}.

A further superexponential, contribution to $\rho(\vol)$ comes from
manifolds with the product topology $M\approx\Sigma_1\times\Sigma_2$.
Any surface of genus $h>1$ admits a moduli space ${\cal M}_h$ of constant
negative curvature metrics \cite{Gardiner}, and a pair of metrics drawn
from ${\cal M}_{h_1}\times{\cal M}_{h_2}$ determines an Einstein
(although not hyperbolic) metric on $\Sigma_1\times\Sigma_2$.  By the
Gauss-Bonnet theorem, the volume of a genus $h$ surface with a hyperbolic
metric is
\beq
{\mathit{Vol}}(\Sigma) = \int\sqrt{g}\,d^2x =
  {1\over2\Lambda}\int R\sqrt{g}\,d^2x =
  {\pi\over\Lambda}\chi(\Sigma) = {2\pi\over\Lambda}(1-h) ,
\label{c3}
\eeq
and hence
\beq
\vol(\Sigma_1\times\Sigma_2) = {4\pi^2\over9}(h_1-1)(h_2-1) .
\label{c4}
\eeq
The number of different product manifolds with normalized volume $V$ is
thus roughly equal to the number of factors of $9V/4\pi^2$, which does
not grow superexponentially.  In contrast to the hyperbolic case, however,
there are many Einstein metrics on each manifold $\Sigma_1\times\Sigma_2$,
and $\rho(\vol)$ must include a factor of the volume of the moduli space
of such metrics.  Now, the moduli space ${\cal M}_h$ for a surface $\Sigma$
has a volume that grows factorially with the genus $h$ \cite{Penner,Gross2}.
Hence the corresponding volume of the moduli space of Einstein metrics on
$\Sigma_1\times\Sigma_2$ grows at least as fast as $h_1!h_2!$.  Product
manifolds thus contribute a term
\beq
\rho(\vol) \sim \sum_{mn=9\vol/4\pi^2} (m+1)!(n+1)!
\sim \exp\left\{ {9{\vol}\over4\pi^2}\log\vol \right\}
\label{c5}
\eeq
to the density of topologies.

\section{Implications and Speculations}

The clearest implication of the preceding analysis is that the sum over
topologies is qualitatively different for $\Lambda>0$ and $\Lambda<0$.
While there may be some manifolds that admit Einstein metrics with both
signs of $\Lambda$, many do not.  In particular, the hyperbolic manifolds
that lead to the factorial growth \rref{c2} in $\rho(\vol)$ do not
contribute at all to the saddle point approximation for $\Lambda>0$.  The
case of vanishing cosmological constant is different still: the classical
action is zero, and the partition function is controlled by the one-loop
determinants.  The value $\Lambda=0$ thus appears to be a sort of boundary
between phases.  This does not in itself explain why $\Lambda$ should vanish,
of course, but it is suggestive.

For $\Lambda\ne0$, at least in the saddle point approximation, the key
issue is the balance between ``action'' and ``entropy'' in the sum
\rref{a8} and similar path integrals.  Since the full behavior of
$\rho(\vol)$ is not known for either sign of $\Lambda$, the conclusions
we can draw are limited.  Nevertheless, there is room for some interesting
speculation.

For $\Lambda<0$, it is evident that the factorial growth \rref{c2} is
sufficient to guarantee that the sum \rref{a8} fails to converge.  To say
more, we need to understand relative phases: if the terms in \rref{a8}
have identical phases, the series is not even Borel summable, but if the
phases differ, it may be possible to define a Borel sum \cite{Borel}.
Relative phases can come from the one-loop determinants $\Delta_M$, or
more precisely from negative-eigenvalue modes \cite{Gross} (and perhaps
the zero-modes \cite{Hawking3}) of the operators whose determinants
appear in this prefactor.  The hyperbolic manifolds considered in the
preceding section have no global symmetries and no moduli spaces, so
no zero-modes are expected.  However, the relevant one-loop operator
$\Delta^\Lambda(1,1)$, given by
\beq
\Delta^\Lambda(1,1) \phi_{\mu\nu} = -\nabla_\rho\nabla^\rho \phi_{\mu\nu}
   - 2R_{\mu\rho\nu\sigma}\phi^{\rho\sigma}
\label{d0}
\eeq
acting on symmetric transverse traceless tensors \cite{Christensen,Gross},
is not positive definite for hyperbolic manifolds, and negative modes
may occur.  As discussed earlier, additional phases may also be introduced
by hand---for example, by adding a term to the action proportional to
the Euler characteristic---so the question of Borel summability remains
unresolved.

It is worth reemphasizing that the factorial growth \rref{c2} is only a
lower bound; I do not know whether it provides a good estimate for the
actual behavior of $\rho(\vol)$.  We may be able to learn more about
this question from lattice formulations of quantum gravity.  In random
triangulation models, the number of geometries on a fixed four-manifold
grows exponentially with the number of four-simplices, while the total
number of geometries on {\em all\/} manifolds grows factorially
\cite{Carfora}.  The geometries counted by these models are not the
same as those occurring in \rref{a8}---the metrics need not be Einstein
metrics---but the conclusions are similar enough to suggest that the
factorial growth \rref{c2} may describe the actual behavior of $\rho(\vol)$,
and not merely provide a lower bound.

A very similar divergence in the sum over topologies occurs in string
theory \cite{Gross2}.  In two dimensions, the divergence can be handled
by appealing to matrix models \cite{matrix}, although the cure requires
that we abandon any fundamental role for smooth geometries.  In four
dimensions, we presently know of no comparable solution, but the sum over
topologies may ultimately be explained as an expansion in some coupling
constant in a more fundamental theory.

Even without such an underlying theory, though, it may be possible to
reach some tentative conclusions about the sum over topologies in quantum
gravity.  The partition function \rref{a8} is formally identical to that of
a thermodynamic system, with $\vol$ serving as ``energy,'' $-\Lambda$ acting
as a ``temperature,'' and $\rho(\vol)$ playing the role of a density of
states.  As discussed in reference \cite{Carlip2}, the superexponential
growth \rref{c2} for $\Lambda<0$ is analogous to the behavior of a system
with negative heat capacity:  just as the microcanonical temperature of
such a system is driven to zero as the energy rises, the ``microcanonical
cosmological constant''
\beq
\Lambda_{\mathrm{micro}} = - {9\over8\pi L_P{}^2}
  \left( {\partial\ln\rho(\vol)\over\partial\vol} \right)^{-1}
\label{d1}
\eeq
goes to zero as $\vol\rightarrow\infty$.  In a thermodynamic system, this
behavior has a straightforward physical origin: rather than increasing
the temperature, the addition of energy drives the creation of new states,
which are produced so copiously that the energy per state falls.  It was
argued in \cite{Carlip2} that the same may be true in quantum gravity,
where processes that would normally increase the absolute value of the
vacuum energy might instead merely drive the production of more and more
complicated spacetime foam.

For $\Lambda>0$, the thermodynamic analogy also seems to work surprisingly
well.  From equation \rref{a8}, a positive cosmological constant  is
analogous to a negative temperature.  Negative temperatures typically occur
in spin systems, which are characterized by a finite number of states
and a maximum energy.  It is not known whether the number of manifolds
admitting an Einstein metric with $\Lambda>0$ is finite, but it is
certainly true that $\vol$, the analog of the energy, has a maximum
value \rref{b1}, and that it takes that value for only a single topology.

In order to decide whether these analogies are more than coincidences,
we need to answer questions like the following: If a phase transition
in matter takes place at time $t_1$, leading to a nonvanishing vacuum
energy density, what is the most likely topology of the universe at time
$t_2>t_1$, and what is the probability that the universe will appear to
have a nonzero cosmological constant at that time?  Questions of this sort
call for a dynamical description of spacetime topology, and such dynamical
accounts are notoriously difficult in quantum gravity, requiring us to
confront the ``problem of time'' \cite{Kuchar}.  For a path integral
approach of the kind presented here, the consistent history approach
to quantum gravity may offer a fruitful avenue for further research
\cite{Hartle1}.

Finally, it is interesting to consider implications of this work for the
Hartle-Hawking wave function of the universe \cite{Hartle2}.  A manifold
with a positive definite Einstein metric is a real tunneling geometry---an
instanton for ``creation of a universe''---if it can be cut along a
codimension-one hypersurface $\Sigma$ of vanishing extrinsic curvature.
The resulting boundary then determines an initial state for a Lorentzian
universe \cite{Hartle3}.  If $\Sigma$ is separating, this cutting process
yields two disjoint pieces, each with a single-component boundary.  If
$\Sigma$ is not separating, the resulting manifold has two boundary
components, and may represent a ``multiple birth'' of disconnected
universes.

In the construction described in section \ref{sec1}, the ``seed manifold''
$M$ always contains a three-dimensional submanifold $\Sigma$ of vanishing
extrinsic curvature \cite{Lubotzky2}, and can thus be viewed as a real
tunneling geometry.  Moreover, $\Sigma$ lifts to $n$ disjoint copies
of itself in each of the $n$-fold covering spaces used in the construction
\cite{Lubotzky1}.  This means that the covering spaces are themselves
real tunneling geometries, each carrying an identical induced metric
$g_{\Sigma}$ on a totally geodesic hypersurface $\Sigma$.  I do not
know whether these hypersurfaces are separating.  If they are, then the
derivation of section \ref{sec1} demonstrates that the Hartle-Hawking
wave function is infinitely peaked at the geometry $g_\Sigma$, much as
it is in 2+1 dimensions \cite{Carlip3}.  If they are not, it may still
be possible to cap off one of the boundaries in the resulting ``multiple
birth'' geometries, in which case there will again be an infinite peak
in the wave function.  Work on this question is in progress.

A similar phenomenon occurs for the product manifolds discussed in
section \ref{sec1}.  For these topologies, a real tunneling geometry can
always be obtained by choosing a metric on one of the two surfaces---say
$\Sigma_1$---that admits an orientation-reversing involution \cite{Gibbons2}.
The resulting Lorentzian universe has the spatial topology
$S^1\times\Sigma_2$, with an initial hypersurface that carries a direct
sum metric $d\theta^2\oplus g_2$ for some metric $g_2$ in the moduli space
${\cal M}_{h_2}$.

The Hartle-Hawking wave function is a functional of this boundary metric,
and can thus be viewed as a function on ${\cal M}_{h_2}$.  If we fix a point
in ${\cal M}_{h_2}$, contributions to the wave function will come from
manifolds $\Sigma_1\times\Sigma_2$ for every topology $\Sigma_1$ of genus
$h_1>1$, and for every metric $g_1\in{\cal M}_{h_1}$ on $\Sigma_1$ that
admits an orientation-reversing involution.  For each $h_1>1$, the
corresponding density of topologies will therefore include a factor
proportional to the volume of ${\cal M}_{h_1}^+$, the moduli space of
metrics that admit orientation-reversing involutions.  While the arguments
of reference \cite{Penner} do not apply directly to this moduli space, it
is very likely that they can be extended to show that its volume grows
superexponentially with $h$.  If this is the case, the Hartle-Hawking
wave function will again have infinite peaks, now for the topologies
$S^1\times\Sigma_2$.

\newpage
\vspace{1.5ex}
\begin{flushleft}
\large\bf Acknowledgements
\end{flushleft}

I received invaluable help in this work from a number of mathematicians,
including Walter Carlip, Greg Kuperberg, Alex Lubotzky, Alex Nabutovsky,
and Bill Thurston.  In particular, the structure of the proof in section
\ref{sec1} was suggested by Bill Thurston and independently by
Alex Lubotzky, and the details were shown to me by Lubotzky.  This
research was supported in part by National Science Foundation grant
PHY-93-57203 and Department of Energy grant DE-FG03-91ER40674.

\end{document}